\documentclass[sigconf]{acmart}

\renewcommand\footnotetextcopyrightpermission[1]{} 

\usepackage{booktabs} 

\setcopyright{none}

\acmDOI{}

\acmISBN{}

\acmConference{KDD 2018}{London}{United Kingdom} 
\acmYear{2018}
\copyrightyear{2018}

\acmPrice{}

\begin{document}
\title{Thresholded ConvNet Ensembles: Neural Networks for Technical Forecasting}

\author{Sid Ghoshal}
\affiliation{%
  \institution{Department of Engineering Science\\
  	Oxford-Man Institute of Quantitative Finance}
  \city{University of Oxford} 
}
\email{sghoshal@robots.ox.ac.uk}

\author{Stephen Roberts}
\affiliation{%
  \institution{Department of Engineering Science\\
  	Oxford-Man Institute of Quantitative Finance}
  \city{University of Oxford} 
}
\email{sjrob@robots.ox.ac.uk}


\begin{abstract}
Much of modern practice in financial forecasting relies on technicals, an umbrella term for several heuristics applying visual pattern recognition to price charts. Despite its ubiquity in financial media, the reliability of its signals remains a contentious and highly subjective form of `domain knowledge'. We investigate the predictive value of patterns in financial time series, applying machine learning and signal processing techniques to 22 years of US equity data. By reframing technical analysis as a poorly specified, arbitrarily preset feature-extractive layer in a deep neural network, we show that better convolutional filters can be learned directly from the data,  and provide visual representations of the features being identified. We find that an ensemble of shallow, thresholded CNNs optimised over different resolutions achieves state-of-the-art performance on this domain, outperforming technical methods while retaining some of their interpretability.
\end{abstract}

%
%



\maketitle

\section{Introduction}

In financial media, extensive attention is given to the study of charts and visual patterns. Known as \textit{technical analysis} or \textit{chartism}, this form of financial analysis relies solely on historical price and volume data to produce forecasts, on the assumption that specific graphical patterns hold predictive information for future asset price fluctuations (Blume et al, 1994). Early research into genetic algorithms devised solely from technical data (as opposed to e.g. fundamentals or sentiment analysis) showed promising results, sustaining the view that there could be substance to the practice (Neely et al, 1997; Allen and Karjailainen, 1999). Research in finance has typically restricted itself to the time series of closing prices and the visuals emerging from line charts (Lo et al, 2000), relying on kernel regression to smooth out the price process and enable pattern recognition. 

An equally common visual representation of price history in finance is the candlestick. Candlesticks encode opening price, closing price, maximum price and minimum price over a discrete time interval, visually represented by a vertical bar with lines extending on either end. Much as with line charts, technical analysts believe that specific sequences of candlesticks reliably foreshadow impending price movements. A wide array of such patterns are commonly watched for (Taylor and Allen, 1992), each with their own pictogram and associated colourful name (`inverted hammer', `abandoned baby', etc).  

Recent research on candlestick chartism has debunked the validity of several highly-cited patterns (Ghoshal and Roberts, 2017). Drawing on a modern intuition for pattern recognition in vision and language (Bengio, 2009), candlestick patterns are reframed as a form of \textit{feature engineering} intended by chartists to extract salient features, facilitating the classification of future returns with higher fidelity than the raw price process would otherwise allow. Filters learned through convolution show promise, and serve as the blueprint for an interpretable application of deep learning to the financial domain.

After defining a format for cross-correlating time series data with chartist filters (Section 2.1-2.3), we undertake a thorough statistical assessment of the predictive prowess of 1-day, 2-day and 3-day candlestick formations (Section 2.4). Feeding candlestick data through a neural network involving separate filters for each technical pattern, we classify next-day returns with the filters implied by chartist doctrine (Section 3.1-3.2) and set this cross-correlational approach as a baseline to improve upon (Romaszko, 2015). We then compare the model's accuracy when filters are not preset but instead learned by thresholded convolutional neural networks (TCNNs) during their training phase (Section 3.3). Finally we assess the significance of our findings (Section 3.4), and benchmark deep learning in finance against alternative methods (Section 3.5).

The contributions of this paper are threefold: firstly, we rigorously evaluate the practice of candlestick chartism, and find little evidence to support it. We agree with Lo et al (2000) that the distribution of future returns conditioned on observing technical patterns diverges significantly from the unconditional distribution, but upon close inspection the resulting classifier barely outperforms guesswork. Secondly, we show that filters learned and tested on 22 years of S\&P500 price data in a CNN architecture can yield modest gains in accuracy over technical methods. Thirdly, we demonstrate that considerable gains in forecasting capability are achievable through ensemble methods and confidence thresholds.

\section{Evaluating Technical Analysis}

\subsection{Definition of Candlestick Data}

Both the financial time series data and the candlestick technical filters used by chartists take the same form. Asset price data for a discrete time interval is represented by four features: the opening price (price at the start of the interval), closing price (price at the end of the interval), high price (maximum over the interval) and low price (minimum over the interval). The candlestick visually encodes this information (Fig. 1): the bar's extremities denote the open and close prices, and the lines protruding from the bar (the candle's `wicks' or shadow) denote the extrema over the interval. The colour of the bar determines the relative ordering of the open and close prices: a white bar denotes a positive return over the interval (close price $>$ open price) and a black or shaded bar denotes a negative return (close price $<$ open price).

\begin{figure}[ht]
	\centering
	\includegraphics[scale=0.6]{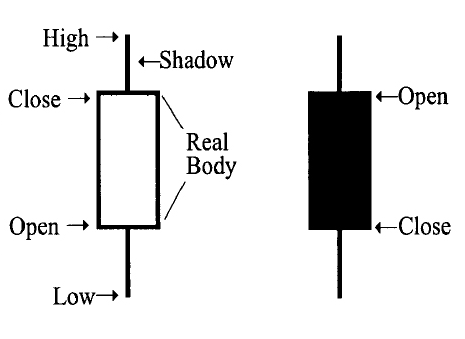}
	\begin{center}
		\footnotesize
		Figure 1: Candlestick representation of financial time series data. 
	\end{center}
\end{figure}

We can therefore summarise the candlestick representation of a financial time series of length $n$ timesteps as a $4\times n$ price signal matrix $F$ capturing its four features. Throughout this paper we rely on daily market data, but the methods can be extended to high-frequency pattern recognition using tick data and full order books.

\subsection{Definitions of Technical Patterns}

We undertake a comprehensive review of all the major candlestick patterns cited by practitioners of technical analysis at multiple timescales. The simple 1-day patterns include the hammer (normal and inverted), hanging man, shooting star, dragonfly doji, gravestone doji, and spinning tops (bullish and bearish). Our 2-day patterns cover the engulfing (bullish and bearish), harami (bullish and bearish), piercing line, cloud cover, tweezer bottom and tweezer top. Finally our 3-day patterns cover some of the most cited cases in chartist practice:  the abandoned baby (bullish and bearish), morning star, evening star, three white soldiers, three black crows, three inside up and three inside down. Fig. 2 provides both the visual template associated with each pattern, as well as the future price direction it is meant to presage. As before, we summarise a technical pattern $P$ of length $m$ timesteps as a $4\times m$ matrix $T_{P_m}$, standardised for comparability to have zero mean and unit variance.


\begin{figure}[ht]
	\centering
	\includegraphics[scale=0.25]{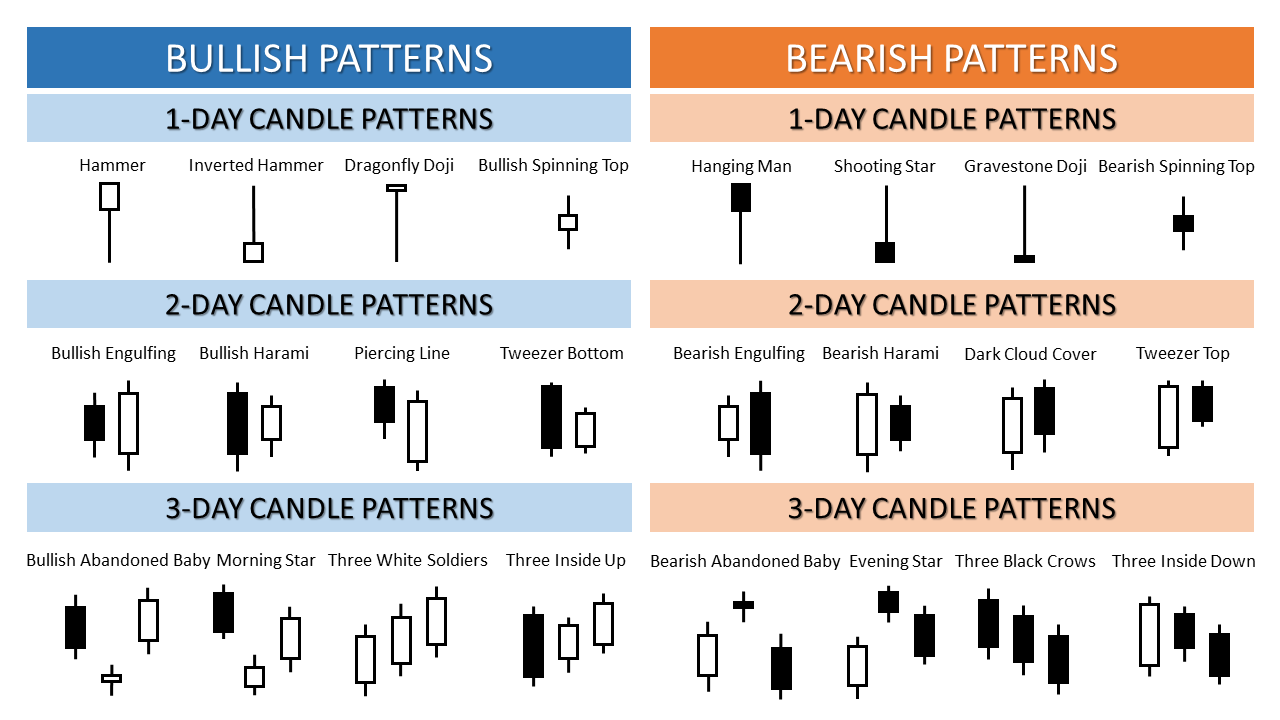}
	\begin{center}
		\footnotesize
		Figure 2: For each timescale (1-day, 2-day and 3-day), we specify 8 chartist patterns and the future direction they predict (`bullish' for positive returns, `bearish' for negative returns).
	\end{center}
\end{figure}

\subsection{Identification by Template Matching}

Matrix representations for both the template $T_{P_m}$ and equal-length, standardised rolling windows $F_{n}$ of the full price signal $F$ at timestep $n$ can be cross-correlated together to generate a time series $S_P$ measuring the degree of similarity between the price signal and the filter. For a given pattern $P$, at each timestep $n$:

\begin{equation}
S_{P,n} = \bigg\langle\frac{T_{P_m}}{\|T_{P_m}\|},\frac{F_{n}}{\|F_{n}\|},\bigg\rangle
\end{equation}

\vskip 10pt

\noindent where $\langle\cdot,\cdot\rangle$ is the inner product of the two matrices and $\|\cdot\|$ is the $L^{2}$ norm.

Our algorithm extracts the top centile of similarity scores $S_P$ as pattern matches and produces a distribution of next-day returns conditional on matching pattern $P$. 

\subsection{Evaluating Technical Analysis}

We run several diagnostics to assess separately the informativeness and predictive prowess of each technical pattern.

\subsubsection{Empirical Data}

Throughout our work, we use technical (i.e. open, close, high and low price) data from the S\&P500 stock market index constituents for the period Jan 1994 - Dec 2015, corresponding to $n$ = 2,817,849 entries of financial data in the price signal $F$. This dataset covers a representative cross-section of US companies across a wide timeframe suitable for learning the patterns, if any, of both expansionary and recessionary periods in the stock market.

\begin{figure}[ht]
	\centering
	\includegraphics[scale=0.43]{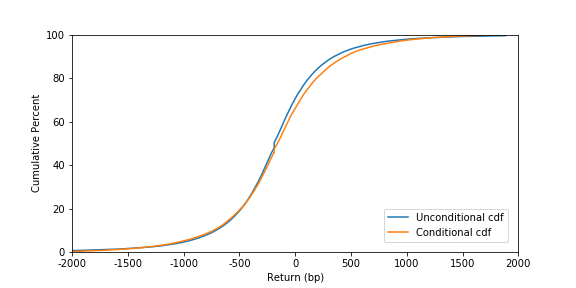}
	\begin{center}
		\footnotesize
		Figure 3: Empirical cumulative distribution functions of unconditional returns and returns conditioned on matching the pattern `Three Black Crows', where a match is deemed to have occurred when the similar score $S_P$ is in its top centile.
	\end{center}
\end{figure}

\begin{figure*}[t]
	\centering
	\includegraphics[scale=0.46]{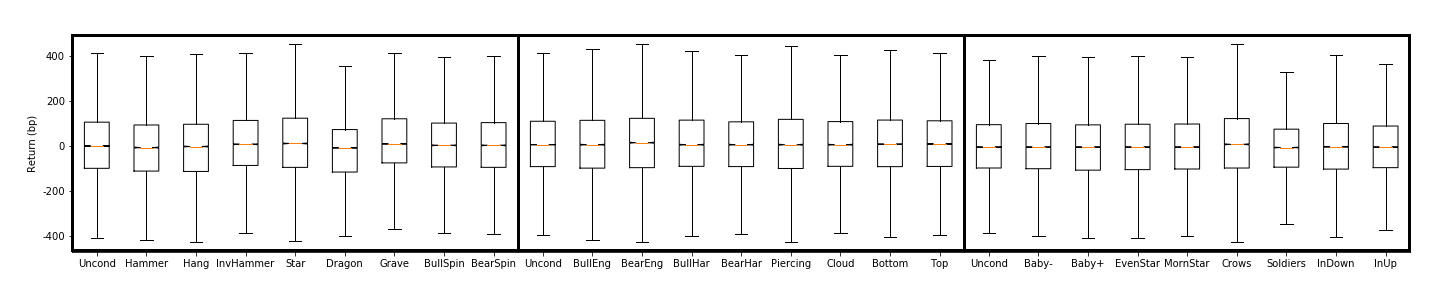}
	\begin{center}
		\footnotesize
		Figure 4: Notched boxplots of the distributions of returns in basis points (one hundredth of a percent), conditional on observing each of the technical patterns (similarity score $S_P$ in its top centile). At a glance, none of the conditional distribution medians diverge substantively from the unconditional baseline, and the distributions' standard deviations dwarf their medians by two orders of magnitude.
	\end{center}
\end{figure*}

\begin{figure*}[t]
	\centering
	\includegraphics[scale=0.46]{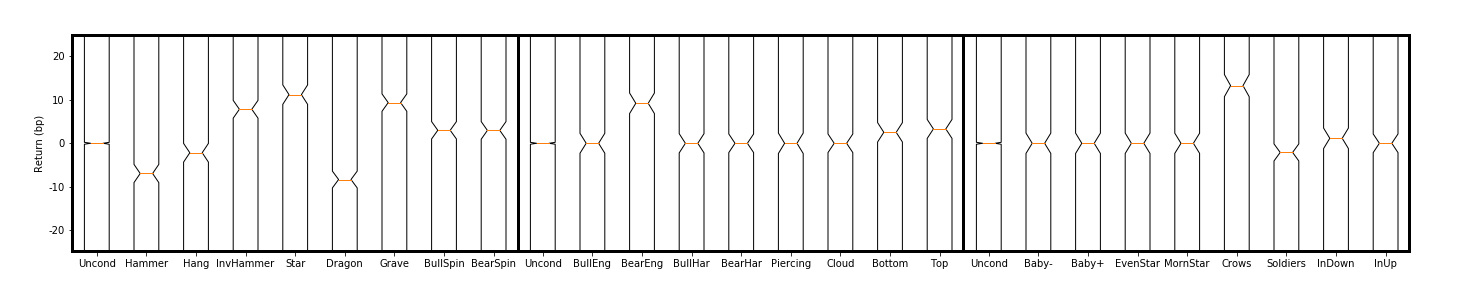}
	\begin{center}
		\footnotesize
		Figure 5: Close-up of boxplot notches for the distributions of returns in basis points (one hundredth of a percent), conditional on observing each of the technical patterns (similarity score $S_P$ in its top centile). Surprisingly, several single-day patterns do in fact correlate with abnormal next-day returns. Almost all of the multi-day patterns exhibit notches that overlap with the unconditional distribution's, implying that the distribution medians are not meaningfully changed by conditioning. Only `Bearish Engulfing' and `Three Black Crows' seem to be significant - as harbingers of better times, despite their names.
	\end{center}
\end{figure*}

\subsubsection{Informativeness}

We begin by comparing the top centile of conditional returns with their unconditional counterparts, with the view that conditioning on informative patterns should yield significantly different distributions. Denoting by $\{{R_P}_{t=1}^{n_1}\}$ the subset of returns conditioned on matching pattern $P$ and $\{{R}_{t=1}^{n_2}\}$ the full set of unconditional returns, we compute their empirical cumulative distribution functions $F_1(z)$ and $F_2(z)$. The two-sample Kolmogorov-Smirnov (K-S) test evaluates the null hypothesis that the distributions generating both samples have identical cdfs, by computing the K-S statistic:

\begin{equation}
\gamma = \bigg(\frac{n_1 n_2}{n_1+n_2}\bigg)^{1/2} \sup_{-\infty<z<\infty} |F_1(z) - F_2(z)|
\end{equation}

\vskip 10pt

The limiting distribution of $\gamma$ provides percentile thresholds above which we reject the null hypothesis. When this occurs, we infer that conditioning on the pattern does materially alter the future returns distribution. As an example of this approach, we provide the empirical cdfs of both unconditional returns and returns conditioned on the pattern: `Three Black Crows' (Fig. 3).

\subsubsection{Predictive Prowess}

Whilst these patterns may bear some information, it does not follow that their information is actionable, or even aligns with the expectations prescribed by technical analysis. Notched boxplots of both unconditional returns and returns conditioned on each of the filters (Fig. 4) allow us to gauge whether the pattern's occurrence does in fact yield significant returns in the intended direction. 

A closer examination suggests several of the 1-day patterns are in fact relevant, but that the more elaborate 2-day and 3-day formations are not. Conditioning on 14 of the 16 multi-day patterns produces no significant alteration in the median of next-day returns distributions (Fig. 5): only the `Bearish Engulfing' and `Three Black Crows' patterns produce a conditional distribution for which the 95\% confidence interval of the median (denoted by the notch) differs markedly from its unconditional counterpart.

\subsubsection{Results}

We report the empirical results of the K-S goodness of fit tests and top centile (Table 1) conditional distribution summary statistics, using daily stock data from the S\&P500. Though several of the patterns do indeed bear information altering the distribution of future returns, their occurrence is neither a reliable predictor of price movements (high standard deviation relative to the mean) nor even, in many instances, an accurate classifier of direction. Elaborate multi-day patterns systematically perform worse than their single-day counterparts. Surprisingly, 6 of the 8 single day patterns do in fact produce meaningful deviations from the unconditional baseline, with the dragonfly and gravestone doji standing out as significant outliers (-25.81 bp and +22.41 bp respectively when conditioning on the top centile of similarity score, Table 1). But even in those instances, technical analysis gets the direction wrong: the deviations are the polar opposite of what chartist doctrine would imply. Conceptually, the notion of using filters in financial data to extract informative feature maps may bear merit - but the chartist filter layer is demonstrably an improper specification.

\begin{table}[ht]
	\caption{Summary statistics for the next-day return distributions conditioned on matching technical patterns. A match on pattern $P$ is deemed to have occurred when the cross-correlational similarity score $S_P$ is in its top \textit{centile}. K-S statistics $\gamma$ above 1.95 are significant at the 0.001 level. Mean return $\mu$ for each pattern is expressed as a difference from the unconditional baseline. The incremental mean returns are dwarfed by their standard deviation, and do not even always move in the direction prescribed by chartism.}
	\begin{center}
		\begin{small}
			\begin{sc}
				\begin{tabular}{lcccr}
					\hline
					Pattern & $\gamma$ & $\mu (bp)$ & $\sigma (bp)$ \\
					\hline
					Unconditional    &  & 4.26 & 229.40 \\
					\hline
					\textcolor{green}{Hammer} & 5.13& \textcolor{red}{-15.80} & 223.04\\
					\textcolor{green}{Inverted Hammer} & 5.00 & \textcolor{green}{+13.75} & 211.62\\
					\textcolor{red}{Hanging Man} & 3.71 & \textcolor{red}{-14.92} & 222.06\\
					\textcolor{red}{Shooting Star} & 4.78& \textcolor{green}{+12.01} & 232.42\\
					\textcolor{green}{Dragonfly Doji} & 14.73& \textcolor{red}{-25.81} & 219.99\\
					\textcolor{red}{Gravestone Doji} & 12.93& \textcolor{green}{+22.41} & 223.57\\
					\textcolor{green}{Bullish Spinning Top} & 2.64& \textcolor{red}{-0.72} & 214.70\\
					\textcolor{red}{Bearish Spinning Top} & 1.67& \textcolor{green}{+0.94} & 213.30\\
					\hline
					\textcolor{green}{Bullish Engulfing} & 1.61& \textcolor{red}{-0.28} & 236.50\\
					\textcolor{red}{Bearish Engulfing} & 4.16& \textcolor{green}{+5.75} & 238.47\\
					\textcolor{green}{Bullish Harami} & 1.07& \textcolor{green}{+5.5} & 222.17\\
					\textcolor{red}{Bearish Harami} & 1.51& \textcolor{red}{-0.96} & 219.43\\
					\textcolor{green}{Piercing Line} & 2.29& \textcolor{green}{+0.78} & 241.42\\
					\textcolor{red}{Cloud Cover} & 1.06& \textcolor{red}{-0.75} & 218.24\\
					\textcolor{green}{Tweezer Bottom} & 1.76& \textcolor{green}{+4.19} & 223.23\\
					\textcolor{red}{Tweezer Top} & 1.22& \textcolor{green}{+2.97} & 221.93\\
					\hline
					\textcolor{red}{Abandoned Baby-} & 3.29& \textcolor{red}{-4.04} & 232.45\\
					\textcolor{green}{Abandoned Baby+}    & 1.27& \textcolor{green}{+2.94} & 232.28\\
					\textcolor{red}{Evening Star}    & 2.89& \textcolor{red}{-0.27} & 231.76\\
					\textcolor{green}{Morning Star}     & 1.80& \textcolor{green}{+2.59} & 231.89\\
					\textcolor{red}{Three Black Crows}      & 6.85& \textcolor{green}{+13.09} & 229.40\\
					\textcolor{green}{Three White Soldiers}      & 6.30& \textcolor{red}{-11.77} & 203.26\\
					\textcolor{red}{Three Inside Down}      & 1.63& \textcolor{green}{+2.72} & 233.12\\
					\textcolor{green}{Three Inside Up}      & 2.50& \textcolor{green}{+0.13} & 220.75\\
					\hline
				\end{tabular}
			\end{sc}
		\end{small}
	\end{center}
\end{table}

\vskip 20pt

\section{Feature Engineering in Finance} 

The approach of searching for informative intermediate feature maps in classification problems has seen widespread success in domains ranging from acoustic signal processing (Hinton et al, 2012) to computer vision (Krizhevsky et al, 2012). Where technical analysis uses filters that are arbitrarily pictographic in nature, we learn layers for feature extraction from data. 

We begin by splitting our S\&P500 time series data into training and test sets corresponding to stock prices from 1994-2004 and 2005-2015  respectively.\footnote{Our classes are defined as `negative return' and `strictly positive return'. As zero return days occur (albeit infrequently) in assets with low denomination, we address the issue of class imbalance by adding Gaussian noise with variance $10^{-6}$ to all returns, evenly spreading the zero return days across both classes. The resulting training set class imbalance is neutral (50.1\% strictly positive return days, 49.9\% negative return days), against a very mild positive skew in the test set (51.2\% strictly positive return days, 48.8\% negative return days).} We evaluate the performance of passing the raw data both with and without chartist filters, and subsequently measure the incremental gain from learning optimal feature maps by convolution. The findings are then benchmarked against widely recognised approaches to time series forecasting including recurrent neural networks, nearest neighbour classifiers, support vector machines (SVM) and random forests. 

\begin{table*}[t]
	\setcounter{table}{1}
	\caption{Details of the architecture for a CNN scanning patterns of length $m$. The number of filters in the convolution layer was deliberately kept low (8) and their dimensions (4 $\times$ $m$) match the technical patterns used in Section 3.2, to enable like-for-like comparability with the technical filter approach.}
	
	\begin{center}
		\begin{small}
			\begin{sc}
				\begin{tabular}{lcccccr}
					\hline
					\# & Layer & Units & Activation Function & Dropout & Filter Shape & Outgoing dimensions\\
					\hline
					0    & Input & - & - & - & - & (input) [4 $\times$ 20] \\
					1    & Convolutional & 8 & ReLU & 0.5 & [4 $\times$ $m$] & [8 $\times$ 20] \\
					3   & FC & 64 & ReLU & 0.5 & - & [64]\\
					4  & FC  & 64 & ReLU & 0.5 & - & [64]\\
					5 &  Softmax  & 2 & - & - & - & (output, 2 classes) [2]\\
					\hline
				\end{tabular}
			\end{sc}
		\end{small}
	\end{center}
\end{table*}

\subsection{Multi-Layer Perceptron}

To address issues of scale and stationarity, we process the original 4 $\times$ $n$ price signal matrix $F$ into a new 80 $\times$ $n$ price signal matrix $F$* where each column is a standardised encoding of 20 business days of price data. This encoding provides 4 weeks of price history, a context or `image' within which neural network filters can scan for the occurrence of patterns and track their temporal evolution. We pass $F$* through a multilayer perceptron (MLP) involving fully-connected hidden layers. Preliminary cross-validation experiments with financial time series determined the network topology required for the model to learn from its training data. Insufficient height (neurons per hidden layer) and depth (number of hidden layers) led to models incapable of learning their training data. We settled on 2 fully-connected layers of 64 neurons with ReLU activation functions, followed by a softmax output layer to classify positive and negative returns. Early stopping during the cross-validation phase determined the length of each experiment: 50 to 100 epochs were optimal to avoid the risk of overfitting. Further regularisation was achieved via the inclusion of dropout (Srivastava, 2014) in the fully-connected layers of the network, limiting the model's propensity towards excessive co-adaptation across layers. A heavily-regularised (dropout $= 0.5$) 2-layer MLP is already able to identify some structure in its data (out-of-sample accuracy of 50.6\% after 100 epochs, Table 3). 

\begin{table}[h]
	\setcounter{table}{2}
	\caption{Accuracy obtained after training a 2-layer MLP through a set number of epochs.}
	\begin{center}
		\begin{small}
			\begin{sc}
				\begin{tabular}{lcccr}
					\hline
					Epochs & In Sample (\%) & Out-of-Sample (\%)\\
					\hline
					1    & 50.3 & 50.3 \\
					5    & 50.7 & 50.4 \\
					10   & 51.4 & 50.5 \\
					50   & 52.2 & 50.6 \\
					100  & 52.5  & 50.6 \\
					\hline
				\end{tabular}
			\end{sc}
		\end{small}
	\end{center}
\end{table}

\subsection{Technically-Filtered MLP}

Reframing technical patterns as pre-learned cross-correlational filters, we consider for each pattern length $m$ the 8 pattern matrices $T_{P_m}$ defined visually in Fig. 2. Each such formation, of form $4 \times m$, is stacked along the depth dimension, producing a $4 \times m \times 8$ tensor $T$ whose inner product with standardised windows of the raw price signal $F$ yields  a new $8 \times n$ input matrix $F_T$,

\begin{equation}
F_T = \big\langle T,F\big\rangle.
\end{equation}

\vskip 10pt

\begin{table}[h]
	\setcounter{table}{3}
	\caption{Accuracy obtained after training a technically-filtered MLP using filters of length $m$ = 1 through a set number of epochs; multi-day filters produced similarly lacklustre results. The technical analysis filters produce feature maps with less discernible structure than the original input.}
	\vskip 0.05in
	\begin{center}
		\begin{small}
			\begin{sc}
				\begin{tabular}{lcccr}
					\hline
					Epochs & In Sample (\%) & Out-of-Sample (\%)\\
					\hline
					1  & 50.0  & 50.2 \\
					5   & 50.2 & 49.8 \\
					10  & 50.2 & 49.7 \\
					50  & 50.4 & 49.9 \\
					100 &  50.5 & 49.8 \\
					\hline
				\end{tabular}
			\end{sc}
		\end{small}
	\end{center}
	\vskip -0.1in
\end{table}

\begin{table*}[t]
	\setcounter{table}{4}
	\caption{Accuracy obtained after training a deep neural network with a single convolution layer learning 1-day, 2-day and 3-day patterns.}
	\begin{center}
		\begin{small}
			\begin{sc}
				\begin{tabular}{c|cc|cc|cc}
					\hline
					\multicolumn{1}{c|}{Filter Length} & \multicolumn{2}{|c|}{1-day} & \multicolumn{2}{|c|}{2-day} & \multicolumn{2}{|c}{3-day} \\
					\hline
					Epochs & In Sample (\%) & Out-of-Sample (\%) & In Sample (\%) & Out-of-Sample (\%) & In Sample (\%) & Out-of-Sample (\%)\\
					\hline
					1  & 50.3  & 50.4 & 50.2  & 50.3& 50.1  & 50.2\\
					5   & 50.6 & 50.3 & 50.7 & 50.4 & 50.5 & 50.3\\
					10  & 50.9 & 50.7 & 51.0 & 50.7 & 51.1 & 50.6 \\
					50  & 51.4 & 51.1 & 51.5 & 50.9 & 51.4 & 51.0 \\
					100 &  51.7 & 51.3 &  51.8 & 51.2 &  51.7 & 51.2 \\
					\hline
				\end{tabular}
			\end{sc}
		\end{small}
	\end{center}
\end{table*}

This new input is the result of cross-correlating the raw price signal $F$ with the technical analysis filter tensor $T$, and can be interpreted as the feature map generated by technical analysis. We now use $F_T$ as the input to the same MLP as before and look for improvements in model forecasts. The results we find are consistent with Section 2: using technical analysis for feature extraction hinders the classifier, slightly degrading model performance (out-of-sample accuracy of 49.8\% after 100 epochs using the 1-day patterns, Table 4).

\subsection{Convolutional Neural Network}

We now deepen the neural network by adding a single convolutional layer with 8 filters to our earlier MLP (architecture detailed in Table 2). Separate experiments are run for convolutional filters of size 4, 8 and 12, corresponding to scanning for 1-day, 2-day and 3-day patterns. Their performance is reported in Table 5.  The CNN finds much greater structure in its training data than the MLP could, and generalises better. Accounting for the size of the test set ($n$ = 1,408,679), the leap from the MLP's out-of-sample accuracy of 50.6\% to the 1-day CNN's out-of-sample accuracy of 51.3\% is highly significant.

\begin{figure}[ht]
	\centering
	\vskip -0.02in
	\includegraphics[scale=0.56]{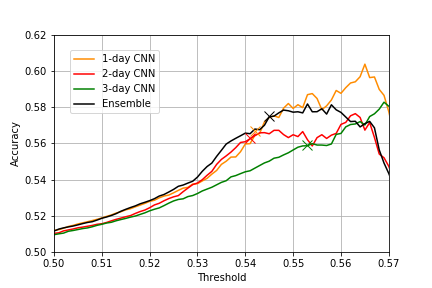}
	\begin{center}
		\footnotesize
		Figure 6: Model accuracy as a function of softmax threshold $\alpha$. For each model, we indicate by a cross the threshold level that retains the 1\% of test data for which the model's output probabilities imply the highest confidence.
	\end{center}
\end{figure}

\subsubsection{Confidence Thresholding}

In contrast to other application domains, finance does not require an algorithmic agent to be accurate at all times. It is acceptable (and factoring in friction costs, preferable) for a model to be sparse in making decisions, only generating `high conviction' calls, if this results in greater accuracy. We replicate this by adding a confidence threshold $\alpha$ to the classification output of the final softmax layer of Table 2: test points where neither class is assigned a probability greater than $\alpha$ are deemed uncertain, and disregarded by the classifier. Accuracy as a function of confidence threshold $\alpha$ is presented in Fig. 6, and demonstrates in all 3 cases that a significant increase in model prowess can be achieved by thresholding the softmax output to only consider class assignments with high certainty. For each model, we also highlight the  $\alpha$ threshold which retains the top centile of test outputs, corresponding to the model's most confident assignments. These vary by model (54.2\%, 54.1\% and 55.3\% for the 1-, 2- and 3-day TCNNs respectively), but in each case form a reliable heuristic for balancing model confidence and sample size. A notable analogue to the study of technical analysis in Section 2:  models searching for more elaborate multi-day patterns tend to underperform the single-day TCNN.

\begin{table*}[t]
	\setcounter{table}{5}
	\caption{Compound Annual Growth Rate (CAGR) and Sharpe ratio of the TCNN models under various assumptions for the cost of trading. The TCNN ensemble and 1-day TCNN are optimal choices for return and risk-adjusted return maximisation, respectively.}
	\begin{center}
		\begin{small}
			\begin{sc}
				\begin{tabular}{c|ccc|ccc|ccc}
					\hline
					\multicolumn{1}{c|}{Friction Cost} & \multicolumn{3}{|c|}{No friction} & \multicolumn{3}{|c|}{0.10\%  per transaction} & \multicolumn{3}{|c}{0.25\%  per transaction} \\
					\hline
					Model & Profit & CAGR (\%) & Sharpe & Profit & CAGR (\%) & Sharpe  &Profit & CAGR (\%) & Sharpe  \\
					\hline
					\textbf{1-day TCNN}  & 46.9  & 42.15 & \textbf{8.34} & 32.8  & 37.72 & \textbf{7.46} & 11.7  & 25.98 & \textbf{5.05}\\
					2-day TCNN   & 36.9 & 39.16 & 8.11 & 22.8 & 33.41 & 6.91 & 1.7 & 9.52 & 1.95 \\
					3-day TCNN  & 44.6 & 41.50 & 6.25 & 30.5 & 36.84 & 5.89 & 9.4& 23.71 & 3.79 \\
					\textbf{TCNN Ensemble}  & \textbf{48.2} & \textbf{42.50} & 6.86 &  \textbf{34.1} & \textbf{38.20} & 6.16 & \textbf{13.0} & \textbf{27.13} & 4.38 \\
					\hline
				\end{tabular}
			\end{sc}
		\end{small}
	\end{center}
\end{table*}

\subsubsection{Ensembling TCNNs}

An effective technique in image processing involves homogeneous ensembling of multiple copies of the same CNN architecture, averaging across the class assignments of the constituent models (Krizhevsky et al, 2012; Antipova et al, 2016). Combining this probabilistic interpretation of the softmax layer with model averaging, we construct a heterogeneous ensemble out of our 1-day, 2-day and 3-day TCNNs. The ensemble benefits from learning patterns manifesting at different timescales, and achieves a higher accuracy (57.5\%) on its top-confidence centile than any of the individual learners (56.7\%, 56.3\% and 55.9\% for the 1-day, 2-day and 3-day TCNN respectively, Fig. 6).

\begin{figure}[ht]
	\centering
	\includegraphics[scale=0.40]{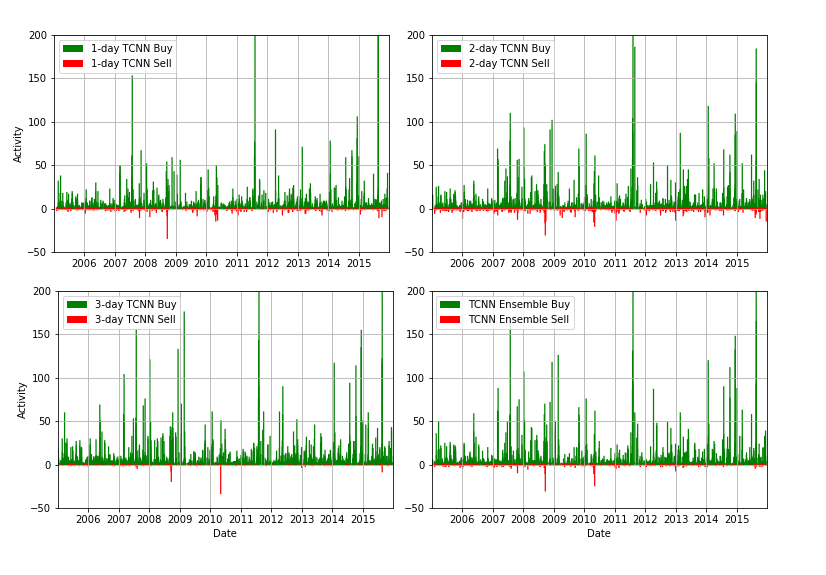}
	\begin{center}
		\footnotesize
		Figure 7: Activity level of the various TCNN models through a 11-year period. As we retain the top centile from the 1,408,679 test points, each model is generating 14,087 trading decisions over 2868 business days, or on average 4.91 trades per day. Though the model is active throughout the window, discernible spikes in activity occur around major events, most notably the US debt ceiling crisis in August 2011. 
	\end{center}
\end{figure}

\subsubsection{Practical Implementation}

Through thresholding, we enforce sparsity in the model's decision making. In a real-world deployment, infrequent activity keeps friction costs low - a desirable outcome for trading algorithms. We track the activity level of the various models over time, as well as the cumulative profit they would generate over the 11-year test window. We assume the model fully captures the 1-day return associated with the top centile of its thresholded class assignments, additively for positive class predictions and subtractively for negative class predictions.

The models are heavily skewed towards buying activity, with accurately-timed spikes centred around major world events (Fig. 7). The 2 largest single-day buy orders occur on the 9th of August 2011 (328 buys), at the tail end of the US debt ceiling crisis which caused the S\&P500 to drop 20\% in 2 weeks, and on the 24th of August 2015 (241 buys), following a flash crash in which US markets erased 12\% of their value before recovering. The largest sell volume occurs on the 22nd of September 2008 (31 sells), a full week after the collapse of Lehman Brothers. This coincides with market-wide relief over Nomura's decision to buy Lehman's operations - and presented the last opportunity to sell before the nosedive of the Great Financial Crisis in late 2008. Despite having no information about world news in their technical dataset, the models were capable of both inferring crucial moments in history, and timing trading decisions around them.


\begin{figure}[ht]
	\centering
	\includegraphics[scale=0.57]{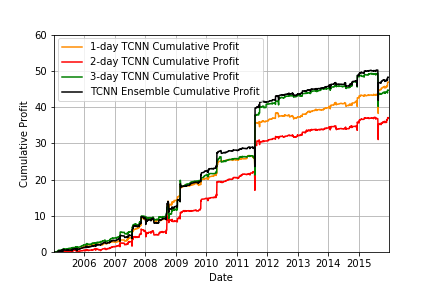}

	\begin{center}
			
		\footnotesize
		Figure 8: Cumulative profit (as a multiple of starting wealth, per Table 6) generated by the various TCNN models between Jan-2005 and Dec-2015, in the absence of friction costs. The models are steadily profitable, with occasional spikes related to recognising major events. Drawdowns are infrequent and of limited scale.
	\end{center}
\end{figure}

\begin{figure*}[t]
	\centering
	\includegraphics[scale=0.61]{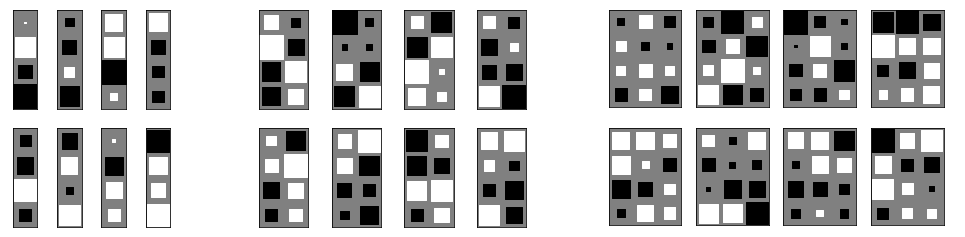}
	\begin{center}
		\footnotesize
		Figure 9: Weight-space visualisation as Hinton diagrams for the 24 cross-correlational filters learned from the first layer of each CNN (8 per constituent model).
	\end{center}
\end{figure*}

Fig. 8 presents the model's profitability over time to highlight the relative steadiness of convolution's performance in identifying stock market patterns, when the decisions are generated by TCNNs and their ensemble. Table 6 translates this performance into compounded annual returns and Sharpe ratios under various assumptions for friction. Even in the absence of tight execution (average trading cost of 0.25\% from the mid-market price), the models remain highly profitable. This sensitivity analysis does nevertheless highlight the importance of good execution in any real-world deployment of algorithmic trading: the TCNN ensemble can only just break even if the per-transaction cost rises to 0.35\%.

\subsubsection{Interpretable Feature Extraction}

\begin{figure}[ht]
	\centering
	\includegraphics[scale=0.5]{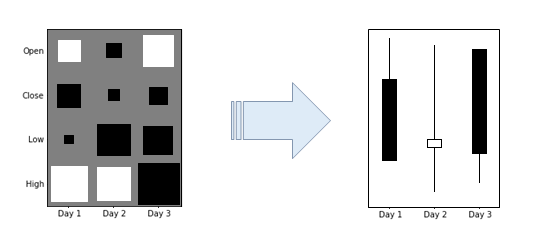}
	\begin{center}
		\footnotesize
		Figure 10: Hinton diagram of the sixth cross-correlational filter learned in the first layer of the 3-day CNN. The relative values of the standardised open, close, low and high for each column in the filter define, in a chartist sense, a specific candlestick sequence (or patch thereof, in instances where the filter's open or close is incompatible with the high-low range) which the neural network extracted as informative for time series forecasting.
	\end{center}
\end{figure}

The convolutional filters learned by the network provide a basis for feature extraction. In particular, the convolutional layer's filters define patches whose cross-correlation with the original input data was informative in minimising both in-sample and out-of-sample categorical cross-entropy. We produce a mosaic of these filters as Hinton diagrams (Fig. 9) and visualise them in the language of technical analysis as candlestick patterns (Fig. 10 and 11), cross-correlational templates whose occurrence is informative for financial time series forecasting. Unlike technical patterns however, these templates have no set meaning: the purpose of individual neurons in a convolutional layer is not readily interpretable.

\subsection{Significance of Model Results}

To investigate whether the predictive performance of the neural network classifiers is statistically significant, we derive the area under the curve (AUC) of each model's receiver operating characteristic curve (ROC), and exploit an equivalence between the AUC and Mann-Whitney-Wilcoxon test statistic $U$ (Mason and Graham, 2002):

\begin{equation}
AUC = \frac{U}{n_P n_N}
\end{equation}

\vskip 10pt

\noindent where $n_P$ and $n_N$ are the number of positive and negative returns in the test set, respectively. In our binary classification setting, the Mann-Whitney-Wilcoxon test evaluates the null hypothesis that a randomly selected value from one sample (e.g., the subset of test data classified as positive next-day returns) is equally likely to be less than or greater than a randomly selected value from the complement sample (the remaining test data, classified as negative next-day returns). Informally, we are testing the null hypothesis that our models have classified at random. $U$ is approximately Gaussian for our sample size, so we compute each model's standardised $Z$-score and look for extreme values that would violate this null hypothesis.

\begin{figure}[ht]
	\centering
	\vskip 5pt
	\includegraphics[scale=0.41]{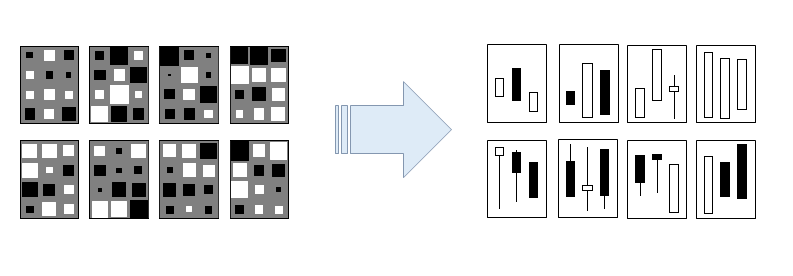}
	\vskip 4pt
	\begin{center}
		\footnotesize
		Figure 11: Candlestick pattern translation of the cross-correlational filter mosaic for the 3-day CNN.
	\end{center}
\end{figure}

\begin{equation}
Z = \frac{U - \mu_U}{\sigma_U}
\end{equation}

\noindent where:

\begin{equation}
\mu_U = \frac{n_P n_N}{2}
\end{equation}

\noindent and

\begin{equation}
\sigma_U = \sqrt{\frac{n_P n_N (n_P + n_N + 1)}{12}}
\end{equation}

\vskip 10pt


\begin{table}[ht]
	\setcounter{table}{6}	
	\caption{AUC, $Z$-score and significance level for the neural network classifiers.}
	\begin{center}
		\begin{small}
			\begin{sc}
				\begin{tabular}{lcccr}
					\hline
					Model & AUC (\%) & $Z$ & Significance\\
					\hline
					MLP  & 51.1 & 23.766 & $>0.9999$\\
					Technical NN  & 49.9 & $-1.878$ & $-$ \\
					1-day CNN  & 51.8 & 36.546 & $>0.9999$ \\
					2-day CNN  & 51.5 & 31.291 & $>0.9999$ \\
					3-day CNN  & 51.5 & 31.423 & $>0.9999$ \\
					CNN Ensemble  & 51.7 & 35.628 & $>0.9999$ \\
					\hline
					1-day TCNN  & 57.2 & 14.533 & $>0.9999$ \\
					2-day TCNN  & 56.5 & 13.017 & $>0.9999$ \\
					3-day TCNN  & 56.2 & 12.493 & $>0.9999$ \\
					TCNN Ensemble  & 57.5 & 15.301 & $>0.9999$ \\
					
					\hline
					
				\end{tabular}
			\end{sc}
		\end{small}
	\end{center}
\end{table}

\noindent Table 7 provides the AUC, $Z$-score and significance of each model, where significance measures the area of the distribution below $Z$. We disregard significance for negative $Z$-scores (as is the case for the technically-filtered neural network) as they imply classifiers that performed (significantly) worse than random chance. Learning neural network filter specifications via convolution yields a significant boost to predictive prowess over the baseline model of Section 3.1 and technically-filtered variant of Section 3.2. Whilst the $Z$-scores of the TCNN models are lower than those of unthresholded CNN models, this is primarily the consequence of sample size on statistical significance tests - AUC improves markedly under thresholding.

\subsection{Performance Benchmarks}

Deep learning has garnered significant attention in recent years for its ability to outperform alternative methods, setting the state-of-the-art in computer vision and speech recognition benchmarks. The lack of commonly-agreed datasets such as MNIST for digit recognition or ImageNet for image classification means finance has lacked a stable backdrop for model benchmarking. For our purposes, we propose the use of the S\&P500 technicals dataset for Jan 1994 - Dec 2015 as a baseline against which to evaluate other classifiers and benchmark deep learning in finance.

\subsubsection{Recurrent Neural Networks (RNN)}

Deep learning for time series analysis has typically relied on recurrent architectures capable of learning temporal relations in the data. Long Short-Term Memory (LSTM) networks have achieved prominence for their ability to memorise patterns across significant spans of time (Hochreiter and Schmidhuber, 1997) by addressing the vanishing gradient problem. A thorough RNN architecture search (Jozefowicz et al, 2015) identified a small but persistent gap in performance between LSTMs and the recently introduced Gated Recurrent Unit (GRU, Chung et al, 2014) on a range of synthetic and real-world datasets. Our benchmark RNNs involve a preliminary recurrent layer (LSTM and GRU, in separate experiments) of 8 neurons followed by 2 dense layers of 64 neurons with dropout, comparable in architectural complexity to the CNN models of Section 3.3.

\subsubsection{k-Nearest Neighbours (k-NN)}

We evaluate a range of nearest neighbour classifiers, labelling each day of the test set with the most frequently observed class label (positive or negative next-day return) in the $k$ training points that were closest in Euclidean space. 

\subsubsection{Support Vector Machines (SVM)}

SVMs have been applied to financial time series forecasting in prior literature, and achieved moderate success when the input features were not raw price data but hand-crafted arithmetic derivations thereof called technical indicators (Kim, 2003). We report SVM performance under different kernel assumptions (linear and RBF), where the model hyperparameters (regularisation parameter $C$ to penalise margin violations, RBF kernel coefficient $\gamma$ to control sensitivity) were selected by cross-validation on a subset of the training data. 

\subsubsection{Random Forests (n-RF)}

In their study of European financial markets, Ballings et al (2015) evaluated the classification accuracy of ensemble methods  against single classifiers. Their empirical work highlighted the effectiveness of random forests in classifying stock price movements and motivates their inclusion in our list of benchmarks, under varying assumptions for the number of trees hyperparameter $n$.

\begin{table}[ht]
	\setcounter{table}{7}	
	\caption{Benchmark performance across a range of supervised learning models trained on S\&P500 technical data for Jan 1994 - Dec 1994 and tested on Jan 2005 - Dec 2015.}
	\begin{center}
		\begin{small}
			\begin{sc}
				\begin{tabular}{lcccr}
					\hline
					Model & Accuracy (\%) & AUC (\%) & $Z$ & Significance\\
					\hline
					MLP  & 50.6 & 51.1 & 23.766 & $>0.9999$\\
					Technical NN  & 49.8 & 49.9 & $-1.878$ & $-$\\
					1-day CNN & 51.3 & 51.8 & 36.546 & $>0.9999$\\
					2-day CNN & 51.2 & 51.5 & 31.291 & $>0.9999$\\
					3-day CNN & 51.2 & 51.5 & 31.423 & $>0.9999$\\							
					CNN Ensemble & 51.2 & 51.7 & 35.628 &$>0.9999$\\
					\hline
					1-day TCNN  & 56.7 & 57.2 & 14.533 & $>0.9999$ \\
					2-day TCNN  & 56.3 & 56.5 & 13.017 & $>0.9999$ \\
					3-day TCNN  & 55.9 & 56.2 & 12.493 & $>0.9999$ \\
					TCNN Ensemble  & 57.5 & 57.5 & 15.301 & $>0.9999$ \\
					
					\hline					
					RNN-LSTM & 50.8 & 51.0 & 19.616 & $>0.9999$\\
					RNN-GRU & 50.9 & 51.2 & 24.880 & $>0.9999$\\
					\hline
					1-NN  & 50.0 & 50.1 & 1.087 & $0.8614$\\
					10-NN  & 49.9 & 49.8 & -3.317 & $-$\\
					100-NN  & 49.7 & 49.6& -7.651 & $-$\\
					\hline					
					Linear SVM & 49.9 & 49.8 & -0.962 & $-$\\
					RBF SVM &  49.9 & 49.8 & -2.416 & $-$\\
					\hline
					10-RF & 50.0 & 50.0 & 0.256 & $0.6009$\\
					50-RF & 49.8 & 49.7 & -5.986 & $-$\\
					100-RF & 49.8 & 49.6 & -7.628 & $-$\\
					\hline

				\end{tabular}
			\end{sc}
		\end{small}
	\end{center}
\end{table}

\subsubsection{Summary}

The results summarised in Table 8 underscore the scale of the challenge for pattern recognition in finance: deep learning achieved the best results by a significant margin, and most alternative methods yielded accuracies that were not statistically distinguishable from guesswork. Convolution outperforms recurrence in our experiments, suggesting that a 20-day window may be sufficient to capture temporal dependencies in markets.

\section{Conclusion}

Our results present to our knowledge the first rigorous statistical evaluation of candlestick patterns in time series analysis, using normalised signal cross-correlation to identify pattern matches. We find little evidence of predictive prowess in any of the standard chartist pictograms, and suspect that the enduring quality of such practices owes much to their subjective and hitherto unverified nature. Nevertheless, it is not inconceivable that price history might contain predictive information, and much of quantitative finance practice relies on elements of technical pattern recognition (e.g., momentum-tracking) for its success. Through a deep learning lens, technical analysis is merely an arbitrary and incorrect specification of the feature-extractive early layers of a neural network. Within relatively shallow architectures, learning more effective filters from data improves accuracy significantly while also providing an interpretable replacement for chartism's visual aids. Thresholding and deep ensembles yield a robust framework for systematic decision making in financial markets, further enhancing performance - though only up to a point. The predictive information embedded in price history appears limited, and even state-of-the-art techniques in pattern recognition will remain subject to that upper bound.

\bibliographystyle{ACM-Reference-Format}

\end{document}